\documentclass{article}

\usepackage{arxiv}

\usepackage[utf8]{inputenc} 
\usepackage[T1]{fontenc}    
\usepackage{hyperref}       
\usepackage{url}            
\usepackage{booktabs}       
\usepackage{amsmath}
\usepackage{amsfonts}       
\usepackage{nicefrac}       
\usepackage{microtype}      
\usepackage{lipsum}
\usepackage{graphicx}
\usepackage{color}
\usepackage{amssymb}
\usepackage{bm}
\usepackage{pgfplots}
\usepackage{tikz}
\usepackage{caption}
\usepackage{subcaption}
\usepackage{multirow}   
\usepackage{algorithm}
\usepackage{algpseudocode}
\usepackage{enumitem}
\usetikzlibrary{arrows,decorations.pathmorphing,backgrounds,positioning,fit,petri, decorations.pathreplacing,shadows,calc}

\pgfplotsset{compat=1.14}

\newcommand{\pluseq}{\mathrel{+}=}
\usepackage{etoolbox}

\newtoggle{redraw}
\newtoggle{redraw2}

\definecolor{voicecolor}{HTML}{BFFFBC}
\definecolor{drumscolor}{HTML}{FFD9CE}
\definecolor{basscolor}{HTML}{A09EBB}
\definecolor{othercolor}{HTML}{CCAD8F}

\definecolor{voicecolordark}{HTML}{60805E}
\definecolor{drumscolordark}{HTML}{806D67}
\definecolor{basscolordark}{HTML}{32313B}
\definecolor{othercolordark}{HTML}{4D4136}

\definecolor{lines}{HTML}{636363}

\definecolor{fmavg}{HTML}{BFFFBC}
\definecolor{largeconvolution}{HTML}{D8FFD6}
\definecolor{interpolation}{HTML}{FFD9CE}
\definecolor{downsample}{HTML}{A09EBB}
\definecolor{regularconvolution}{HTML}{CCAD8F}

\definecolor{voicecolor}{HTML}{BFFFBC}
\definecolor{drumscolor}{HTML}{FFD9CE}
\definecolor{basscolor}{HTML}{A09EBB}
\definecolor{othercolor}{HTML}{CCAD8F}

\tikzset{%
  cascaded/.style = {%
    general shadow = {%
      shadow scale = 1,
      shadow xshift = 9ex,
      shadow yshift = 9ex,
      draw=black,
      thick,
      fill = white},
    general shadow = {%
      shadow scale = 1,
      shadow xshift = 6ex,
      shadow yshift = 6ex,
      draw=black,
      thick,
      fill = white},
    general shadow = {%
      shadow scale = 1,
      shadow xshift = 3ex,
      shadow yshift = 3ex,
      draw=black,
      thick,
      fill = white},
    fill = white, 
    draw,
    thick}}
    
\tikzset{%
pics/rectangle/.style args={#1/#2/#3/#4}{code={%
	\begin{scope}[line width=#2mm]
	\begin{scope}
	\clip (-#1,-#2,0) -- (#1,-#2,0) -- (#1,#2,0) -- (-#1,#2,0) -- cycle;
	\filldraw (-#1,-#2,0) -- (#1,-#2,0) -- (#1,#2,0) -- (-#1,#2,0) -- cycle;
	\end{scope}

	\node[inner sep=0] (-A) at (-#1, 0, #3*0.5) {};
	\node[inner sep=0] (-B) at (#1, 0, #3*0.5) {};
	\node[inner sep=0] (-D) at (0, -0.28, 0.35) {};
	\node[inner sep=0] (-T) at (0, 0.56, 0.35) {};
	
	\coordinate (-V) at (#1, #2);
	\coordinate (-W) at (#1, -#2);
	\end{scope}
}}}

\tikzset{%
  cascadeddrums/.style = {%
    general shadow = {%
      shadow scale = 1,
      shadow xshift = -2ex,
      shadow yshift = 2ex,
      draw=drumscolordark,
      thick,
      fill = drumscolor},
    general shadow = {%
      shadow scale = 1,
      shadow xshift = -1.5ex,
      shadow yshift = 1.5ex,
      draw=drumscolordark,
      thick,
      fill = drumscolor},
    general shadow = {%
      shadow scale = 1,
      shadow xshift = -1ex,
      shadow yshift = 1ex,
      draw=drumscolordark,
      thick,
      fill = drumscolor},
    general shadow = {%
      shadow scale = 1,
      shadow xshift = -.5ex,
      shadow yshift = .5ex,
      draw=drumscolordark,
      thick,
      fill = drumscolor},
    draw=drumscolordark,
    fill = drumscolor, 
    thick}}
    
\tikzset{%
  cascadedother/.style = {%
    general shadow = {%
      shadow scale = 1,
      shadow xshift = -2ex,
      shadow yshift = 2ex,
      draw=othercolordark,
      thick,
      fill = othercolor},
    general shadow = {%
      shadow scale = 1,
      shadow xshift = -1.5ex,
      shadow yshift = 1.5ex,
      draw=othercolordark,
      thick,
      fill = othercolor},
    general shadow = {%
      shadow scale = 1,
      shadow xshift = -1ex,
      shadow yshift = 1ex,
      draw=othercolordark,
      thick,
      fill = othercolor},
    general shadow = {%
      shadow scale = 1,
      shadow xshift = -.5ex,
      shadow yshift = .5ex,
      draw=othercolordark,
      thick,
      fill = othercolor},
    draw=othercolordark,
    fill = othercolor, 
    thick}}
    
\tikzset{%
  cascadedbass/.style = {%
    general shadow = {%
      shadow scale = 1,
      shadow xshift = -2ex,
      shadow yshift = 2ex,
      draw=basscolordark,
      thick,
      fill = basscolor},
    general shadow = {%
      shadow scale = 1,
      shadow xshift = -1.5ex,
      shadow yshift = 1.5ex,
      draw=basscolordark,
      thick,
      fill = basscolor},
    general shadow = {%
      shadow scale = 1,
      shadow xshift = -1ex,
      shadow yshift = 1ex,
      draw=basscolordark,
      thick,
      fill = basscolor},
    general shadow = {%
      shadow scale = 1,
      shadow xshift = -.5ex,
      shadow yshift = .5ex,
      draw=basscolordark,
      thick,
      fill = basscolor},
    draw=basscolordark,
    fill = basscolor, 
    thick}}
    
 \tikzset{%
  cascadedvoice/.style = {%
    general shadow = {%
      shadow scale = 1,
      shadow xshift = -2ex,
      shadow yshift = 2ex,
      draw=voicecolordark,
      thick,
      fill = voicecolor},
    general shadow = {%
      shadow scale = 1,
      shadow xshift = -1.5ex,
      shadow yshift = 1.5ex,
      draw=voicecolordark,
      thick,
      fill = voicecolor},
    general shadow = {%
      shadow scale = 1,
      shadow xshift = -1ex,
      shadow yshift = 1ex,
      draw=voicecolordark,
      thick,
      fill = voicecolor},
    general shadow = {%
      shadow scale = 1,
      shadow xshift = -.5ex,
      shadow yshift = .5ex,
      draw=voicecolordark,
      thick,
      fill = voicecolor},
    fill = voicecolor,
    draw=voicecolordark,
    thick}}
    
\tikzset{%
  cascadeddd/.style = {%
  	general shadow = {%
      shadow scale = 1,
      shadow xshift = -9ex,
      shadow yshift = 9ex,
      draw=voicecolordark,
      thick,
      fill = voicecolor},
    general shadow = {%
      shadow scale = 1,
      shadow xshift = -8.5ex,
      shadow yshift = 8.5ex,
      draw=voicecolordark,
      thick,
      fill = voicecolor},
    general shadow = {%
      shadow scale = 1,
      shadow xshift = -8ex,
      shadow yshift = 8ex,
      draw=voicecolordark,
      thick,
      fill = voicecolor},
    general shadow = {%
      shadow scale = 1,
      shadow xshift = -7.5ex,
      shadow yshift = 7.5ex,
      draw=voicecolordark,
      thick,
      fill = voicecolor},
    general shadow = {%
      shadow scale = 1,
      shadow xshift = -7ex,
      shadow yshift = 7ex,
      draw=drumscolordark,
      thick,
      fill = drumscolor},
    general shadow = {%
      shadow scale = 1,
      shadow xshift = -6.5ex,
      shadow yshift = 6.5ex,
      draw=drumscolordark,
      thick,
      fill = drumscolor},
    general shadow = {%
      shadow scale = 1,
      shadow xshift = -6ex,
      shadow yshift = 6ex,
      draw=drumscolordark,
      thick,
      fill = drumscolor},
    general shadow = {%
      shadow scale = 1,
      shadow xshift = -5.5ex,
      shadow yshift = 5.5ex,
      draw=drumscolordark,
      thick,
      fill = drumscolor},
    general shadow = {%
      shadow scale = 1,
      shadow xshift = -5ex,
      shadow yshift = 5ex,
      draw=drumscolordark,
      thick,
      fill = drumscolor},
    general shadow = {%
      shadow scale = 1,
      shadow xshift = -4.5ex,
      shadow yshift = 4.5ex,
      draw=basscolordark,
      thick,
      fill = basscolor},
    general shadow = {%
      shadow scale = 1,
      shadow xshift = -4ex,
      shadow yshift = 4ex,
      draw=basscolordark,
      thick,
      fill = basscolor},
    general shadow = {%
      shadow scale = 1,
      shadow xshift = -3.5ex,
      shadow yshift = 3.5ex,
      draw=basscolordark,
      thick,
      fill = basscolor},
    general shadow = {%
      shadow scale = 1,
      shadow xshift = -3ex,
      shadow yshift = 3ex,
      draw=basscolordark,
      thick,
      fill = basscolor},
    general shadow = {%
      shadow scale = 1,
      shadow xshift = -2.5ex,
      shadow yshift = 2.5ex,
      draw=basscolordark,
      thick,
      fill = basscolor},
    general shadow = {%
      shadow scale = 1,
      shadow xshift = -2ex,
      shadow yshift = 2ex,
      draw=othercolordark,
      thick,
      fill = othercolor},
    general shadow = {%
      shadow scale = 1,
      shadow xshift = -1.5ex,
      shadow yshift = 1.5ex,
      draw=othercolordark,
      thick,
      fill = othercolor},
    general shadow = {%
      shadow scale = 1,
      shadow xshift = -1ex,
      shadow yshift = 1ex,
      draw=othercolordark,
      thick,
      fill = othercolor},
    general shadow = {%
      shadow scale = 1,
      shadow xshift = -.5ex,
      shadow yshift = .5ex,
      draw=othercolordark,
      thick,
    fill = othercolor},
    fill = white, 
    draw,
    thick}}

\title{Interleaved Multitask Learning for Audio Source Separation with Independent Databases}

\author{
  Clement S. J. Doire\thanks{Equal contribution.}, \, Olumide Okubadejo\textsuperscript{\textasteriskcentered}\\
  Audionamix, Paris\\
}

\begin{document}
\maketitle

\begin{abstract}
Deep Neural Network-based source separation methods usually train independent models to optimize for the separation of individual sources. Although this can lead to good performance for well-defined targets, it can also be computationally expensive. The multitask alternative of a single network jointly optimizing for all targets simultaneously usually requires the availability of all target sources for each input. This requirement hampers the ability to create large training databases. In this paper, we present a model that decomposes the learnable parameters into a shared parametric model (encoder) and independent components (decoders) specific to each source. We propose an interleaved training procedure that optimizes the sub-task decoders independently and thus does not require each sample to possess a ground truth for all of its composing sources. Experimental results on MUSDB18 with the proposed method show comparable performance to independently trained models, with less trainable parameters, more efficient inference, and an encoder transferable to future target objectives. The results also show that using the proposed interleaved training procedure leads to better Source-to-Interference energy ratios when compared to the simultaneous optimization of all training objectives, even when all composing sources are available. 
\end{abstract}

\keywords{Audio Source Separation \and Shared Encoder \and Interleaved Training}

\section{Introduction}
\label{sec:introduction}

Audio source separation has been a key research topic in the audio signal processing community for many years. This is due to its wide-ranging applicability and many practical use cases; from automatic music transcription and lyrics alignment to the spatialization and re-mixing of old recordings. It has attracted even more attention in recent years, in part due to improved result quality brought by deep learning-based methods \cite{SiSEC18}.

Many successful source separation algorithms have been based on Non-negative Matrix Factorization (NMF) \cite{ozerov2010, fevotte2018}. Separating superimposed sources using NMF involves factorizing power spectrograms into the product of two matrices: a dictionary matrix of recurring spectral patterns and its corresponding activation matrix. Although interesting results can be obtained by such methods, for instance when the separation is informed \cite{virtanen2008, hennequin2014}, algorithms based on deep neural networks have been found to deliver better performance \cite{weninger2014}.

Fully-connected feed-forward networks have been applied successfully to source separation by predicting the separated sources spectra one frame at a time and stacking the neighboring frames together to account for temporal context \cite{weninger2014}. Recurrent Neural Networks (RNNs) based on Long-Short-Term-Memory (LSTM) layers \cite{hochreiter1997} have been proposed to better model long term temporal dependencies, leading to significantly better results in source separation \cite{weninger2014, uhlich2017}. Also based on an RNN framework, joint optimization of the source and background masks was proposed in \cite{huang2014}, with an additional discriminative training objective aimed at reducing the interference of the other target sources in the prediction. In \cite{mimilakis2018}, a skip-filtering connection is used in order to predict a time-frequency soft mask while still using the target spectrogram in the computation of the objective function, thus removing the need to explicitly define the ideal target mask.

Convolutional Neural Networks (CNNs) \cite{lecun1998}, originally used for computer vision tasks such as image recognition \cite{simonyan2014}, have also been employed extensively for source separation \cite{chandna2017, jansson2017, takahashi2017, park2018}. They usually present an encoder-decoder structure where, from the input spectrogram, the encoder takes care of successively computing feature maps and downsampling them, while the decoder progressively upsamples the feature maps back to the dimensionality of the original space. Such a bottleneck network structure allows the receptive field of the computed feature maps to increase with depth. This implies the layers can learn meaningful features at different time and frequency scales. In the U-Net architecture \cite{jansson2017}, the downsampling operation is performed using strided convolution layers, and feature maps computed on the downsampling side are concatenated with the feature maps at the same resolution on the upsampling side. This allows more layer interactions as well as reusing previously computed features without loss of information from the successive downsampling operations. A one-dimensional adaptation of the U-Net architecture working directly on the time-domain samples was also proposed in \cite{stoller2018}, in which a learned interpolation layer followed by a unit-stride convolution is used at each upsampling stage to improve on the original transposed convolution. In the MMDenseNet \cite{takahashi2017}, a multi-band structure is proposed where spectrogram bands (e.g. low and high frequency regions) are processed by separate CNNs along with a full-band one before combining their results through a final dense block. Independent networks are trained individually for each target instrument to separate. This work is extended in \cite{takahashi2018} where the MMDenseNet architecture is combined with LSTM blocks, leading to improved performance both with and without additional data. In \cite{park2018}, reduced hourglass modules, close in structure to the U-Net, are stacked sequentially to estimate all the target sources in a single network.

The idea of optimizing for all target sources with a shared network architecture seems promising. Indeed, on top of sharing the computational cost of separating multiple sources in one forward pass, it has been argued that multitask learning can improve generalization by using the information gained from optimizing for related tasks as inductive bias \cite{Caruana1997}. On the other hand, multitask networks can be hard to train as imbalances in the optimized tasks can forbid proper training \cite{chen2018}. To alleviate this problem, dynamic tuning of the gradient magnitudes to balance the individual sub-task losses is proposed in \cite{chen2018}. Taking a continual learning approach, related tasks can also be optimized in a sequential way using elastic weight consolidation to avoid catastrophic forgetting, as proposed in \cite{kirkpatrick2017}. The multitask learning concept was applied to joint source separation and voice activity detection  in \cite{stoller2018b}, leading to improved performance on near-silent regions.

It has been hypothesized in \cite{SiSEC18} that more training data could bring as much improvement in a model's performance as a change in the model itself. Apart from data augmentation techniques \cite{uhlich2017}, one way to greatly improve the amount of training data available is to remove the constraint of having all the target sources available for each input mixture. The resulting independent databases could have, in principle, no data overlap between the different sources to separate. Although using independent databases does not imply any changes when training a separate network for each source separation task, they forbid the training of single network architectures optimizing for all target sources simultaneously. 

In this paper, we investigate the separation of a stereo musical mix into multiple sources as the joint optimization of multiple learning tasks with a shared component, using independent databases. To do so, we use a convolutional neural network composed of a shared encoder for the feature extraction part of the network and independent decoders, one for each of the sources to predict, to handle the non-overlapping sub-tasks. The focus of the present work rests mainly on how to train such a network with independent databases, interleaving mini-batches of data from each individual sub-task, in order to benefit from the potentially greater amount of training data while keeping the advantages of multitask learning. We also investigate the benefit of having part of the network shared among tasks for the estimation of ill-defined targets such as the "other" stem in music source separation. Experimental results show comparable performance between the proposed interleaved training procedure and independently trained models, with less trainable parameters and more efficient inference. The results also show that using the interleaved training procedure leads to higher Source-to-Interference energy ratios when compared to the simultaneous optimization of all training objectives combined in a linear fashion. Finally, we show that slightly increasing the number of feature maps computed at the topmost layers of the shared encoder can have a substantial effect on performance.

The paper is organized as follows. The audio source separation problem as well as the neural network architecture with shared encoder used throughout this paper are detailed in Section~\ref{sec:shared_encoder}. The details of the training procedure with independent databases are given in Section~\ref{sec:training_procedure}. Experimental results are shown and discussed in Section~\ref{sec:evaluation} before concluding in Section~\ref{sec:conclusion}.

\section{Shared Encoder Model}
\label{sec:shared_encoder}

\subsection{Problem Statement}
\label{ssec:problem_statement}

Let $y_{c}(n)$ be the observed mixture signal at discrete-time $n$ for input channel $c \in [1,..,C]$. We compute the magnitude of the complex Short-Time Fourier Transform (STFT) of the observed signal as:
\begin{equation}
    Y_{c}(l,k)=\left| \sum_{n=0}^{K-1}y_{c}(n + lT)w(n)e^{-j\frac{2\pi}{K}nk} \right|
\end{equation}
where $l \in [1,.., L]$ and $k \in [1,.., K]$ are the STFT time-frame index and frequency bin respectively, $w(n)$ a time-domain window and $T$ the frame increment.

Stacking the mixture spectrograms of each input channel, we obtain the observed tensor $\bm{Y}$ of size $(C, L, K)$. We therefore have:
\begin{equation}
    \bm{Y}=f\left(\bm{S}_1, ... , \bm{S}_M\right)
\end{equation}
in which  $f$ is the mixing model and $\bm{S}_i$ for $i=1...M$ are the magnitude spectra of the sources to estimate. In the remainder of the paper, we study the task of separating stereo mixture signals into four components: vocals, drums, bass and other (i.e. the remaining music) denoted by $\bm{V}$, $\bm{D}$, $\bm{B}$ and $\bm{O}$ respectively. We wish to find $\Phi$ such that 
\begin{equation}
\hat{\bm{V}},\, \hat{\bm{D}},\, \hat{\bm{B}},\, \hat{\bm{O}} = \Phi\left(\bm{Y}\right)
\end{equation}
where estimates are denoted by $\,\hat{}\,$ and $\Phi$ is implemented via a neural network. The time-domain signals of the separated sources are then reconstructed using the inverse STFT with the estimated magnitude spectra and the phase of the input mixture signal.

In order to separate the learning of a shared representation from the learning of individual representations embedded in each sub-task, we decompose the parameterized ensemble model $\Phi$ into two sub-models. A shared encoder model $\mathcal{E}$ and task specific decoder models $\mathcal{D}_V,\, \mathcal{D}_D,\, \mathcal{D}_B,\, \mathcal{D}_O$. The model for each sub-task of estimating source $\bm{S}_i$ thus becomes
\begin{equation}
     \hat{\bm{S}}_i = \mathcal{D}_{S_i}\left( \mathcal{E}\left(\bm{Y};\theta_{\mathcal{E}}\right);\theta_{\mathcal{D}_{S_i}}\right)
\end{equation} where model parameters are denoted by $\theta$ with the corresponding model as a suffix and $\bm{S}_{i} \in \mathcal{S} = \left\{\bm{V}, \bm{D}, \bm{B}, \bm{O}\right\}$ the set of target sub-tasks. We use the $L_{1}$ distance as the spectrogram reconstruction loss function during training so that
\begin{equation}
    \label{eq:l1_loss}
    \mathcal{L}_{S_i} = \frac{1}{CLK}\sum_{c=1}^{C}\sum_{l=1}^{L}\sum_{k=1}^{K}\left\Vert \bm{S}_i - \hat{\bm{S}_i} \right\Vert_{1}.
\end{equation}

\subsection{Architecture details}
\label{ssec:architecture_details}

The architecture used throughout this paper takes inspiration from both the U-Net architecture described in \cite{jansson2017} and the reduced hourglass module described in \cite{park2018}. Please note the main focus of this paper is on the concept of using a shared encoder and multiple decoders as well as how to train such an architecture. The same concept can be applied to other network models.

The encoder is akin in structure to that of \cite{park2018}, in which a downsampling stage is alternated with a unit-stride convolution. However, like the U-Net \cite{jansson2017}, we use strided convolution kernels for the downsampling operations. All convolution kernels are followed by Batch Normalization \cite{ioffe2015} and a Leaky Rectified Linear Unit (LeakyReLU) \cite{maas2013} activation function with negative slope $\alpha=0.1$. We use five downsampling stages, doubling the number of computed feature maps at each stage, starting from 16 and stopping at 256.

The task specific decoders all follow the same architecture based on alternating an upsampling stage followed by concatenating the upsampled feature maps with those computed at the same resolution on the encoder side. Following \cite{odena2016} and similar to \cite{park2018, stoller2018}, an upsampling stage consists of nearest-neighbour (NN) interpolation followed by a unit-stride convolution kernel. Similar to the encoder side, each convolution layer is followed by Batch Normalization and a LeakyReLU activation. The number of computed feature maps is regularly halved as resolution increases back to the full spectrogram. Finally, at the end of each decoder, a feature map averaging layer is used followed by a ReLU activation to constrain the estimated magnitude spectrograms to be non-negative. Contrary to \cite{jansson2017, mimilakis2018}, we did not use a soft masking objective as we found it did not improve results in practice. The architecture details and diagram are given in Table~\ref{tab:architecture_details} and Figure~\ref{fig:model_details}.

\begin{figure}
    \centering
    \begin{minipage}{.485\textwidth}
        \vspace{1.69cm}
        \scalebox{0.33}{
        \begin{tikzpicture}
 	\node[] (o2) {};
	\pic[fill=largeconvolution] (O2) {rectangle={4.0/0.5/0.3/1}};
	
	 \node[above=5em of o2, xshift=-.38em] (input) {\hspace{0.4em}\bf{\Huge $\bm{Y}$}};
	
  	\node[below=5em of o2] (downsample1) {};
  	
  	\pic[below=5em of o2, fill=downsample] (Downsample1) {rectangle={4.0/0.5/0.3/1}};
  	
  	\node[below=5em of downsample1] (encoder_conv1) {};
  	
  	\pic[below=5em of downsample1, fill=regularconvolution] (Encoder_conv1) {rectangle={3.5/0.5/0.3/1}};
  	
  	\node[below=5em of encoder_conv1] (downsample2) {};
  	
  	\pic[below=5em of encoder_conv1, fill=downsample] (Downsample2) {rectangle={3.5/0.5/0.3/1}};
  	
  	\node[below=5em of downsample2] (encoder_conv2) {};
  	
  	\pic[below=5em of downsample2, fill=regularconvolution] (Encoder_conv2) {rectangle={3.0/0.5/0.3/1}};
  	
  	\node[below=5em of encoder_conv2] (downsample3) {};
  	
  	\pic[below=5em of encoder_conv2, fill=downsample] (Downsample3) {rectangle={3.0/0.5/0.3/1}};
  	
  	\node[below=5em of downsample3] (encoder_conv3) {};
  	
  	\pic[below=5em of downsample3, fill=regularconvolution] (Encoder_conv3) {rectangle={2.5/0.5/0.3/1}};
 
  	\node[below=5em of encoder_conv3] (downsample4) {};
  	
  	\pic[below=5em of encoder_conv3, fill=downsample] (Downsample4) {rectangle={2.5/0.5/0.3/1}};
  	
  	\node[below=5em of downsample4] (encoder_conv4) {};
  	
  	\pic[below=5em of downsample4, fill=regularconvolution] (Encoder_conv4) {rectangle={2.0/0.5/0.3/1}};
 
  	\node[below =5em of encoder_conv4] (downsample5) {};
  	
  	\pic[below=5em  of encoder_conv4, fill=downsample] (Downsample5) {rectangle={2.0/0.5/0.3/1}};
  	
  	\node[below right=5em and 20em of downsample5] (encoder_conv5) {};
  	
  	\pic[below right =5em and 10em of downsample5, fill=regularconvolution] (Encoder_conv5) {rectangle={1.75/0.5/0.3/1}};

	\node (2) [cascaded, thick,above right=-2em and 2em of encoder_conv5, yshift=-0.7em, minimum width=30em, minimum height=69em, very thick, rectangle, rounded corners] {};

  	\node[right=17em of encoder_conv5] (upsample1) {};
  	
  	\pic[right= 17em of encoder_conv5, fill=interpolation] (Upsample1) {rectangle={1.75/0.5/0.3/1}};
  	
  	\node[above=5em of upsample1, xshift=-0.28em] (decoder_conv1) {};
  	
  	\pic[above=5em of upsample1, xshift=-0.28em, fill=regularconvolution] (Decoder_conv1) {rectangle={2.0/0.5/0.3/1}};
 
  	\node[above=5em of decoder_conv1] (upsample2) {};
  	
  	\pic[above=5em  of decoder_conv1, fill=interpolation] (Upsample2) {rectangle={2.0/0.5/0.3/1}};
  	
  	\node[above=5em of upsample2] (decoder_conv2) {};
  	
  	\pic[above=5em of upsample2, fill=regularconvolution] (Decoder_conv2) {rectangle={2.5/0.5/0.3/1}};
  	
  	\node[above=5em of decoder_conv2] (upsample3) {};
  	
  	\pic[above=5em  of decoder_conv2, fill=interpolation] (Upsample3) {rectangle={2.5/0.5/0.3/1}};
  	
  	\node[above=5em of upsample3] (decoder_conv3) {};
  	
  	\pic[above=5em of upsample3, fill=regularconvolution] (Decoder_conv3) {rectangle={3.0/0.5/0.3/1}};
  
  	\node[above=5em of decoder_conv3] (upsample4) {};
  	
  	\pic[above=5em  of decoder_conv3, fill=interpolation] (Upsample4) {rectangle={3.0/0.5/0.3/1}};
  	
  	\node[above=5em of upsample4] (decoder_conv4) {};
  	
  	\pic[above=5em of upsample4, fill=regularconvolution] (Decoder_conv4) {rectangle={3.5/0.5/0.3/1}};
  	\node[above=5em of decoder_conv4] (upsample5) {};
  	
  	\pic[above=5em  of decoder_conv4, fill=interpolation] (Upsample5) {rectangle={3.5/0.5/0.3/1}};
  	
  	\node[above=5em of upsample5] (decoder_conv5) {};
  	
  	\pic[above=5em of upsample5, fill=regularconvolution] (Decoder_conv5) {rectangle={4.0/0.5/0.3/1}};
  	
  	\node[above=5em of decoder_conv5] (decoder_conv6) {};
  	
  	\pic[above=5em  of decoder_conv5, fill=fmavg] (Decoder_conv6) {rectangle={4.0/0.5/0.3/1}};
  	
 	\node[above=5em of decoder_conv6, xshift=-.37em] (ouput) {\hspace{0.1em}\bf{\Huge $\bm{\hat{S}_i}$}};
	\pic[fill=largeconvolution] (O2) {rectangle={4.0/0.5/0.3/1}};
  	
  	\color{black}

  	  \draw [-stealth, ultra thick] (input) -- node[above] {}  (O2-T);
	\draw [-stealth, ultra thick] (O2-D) -- node[above] {}  (Downsample1-T);
	\draw [-stealth, ultra thick] (Downsample1-D) -- node[above] {}  (Encoder_conv1-T);
	\draw [-stealth, ultra thick] (Encoder_conv1-D) -- node[above] {}  (Downsample2-T);
	\draw [-stealth, ultra thick] (Downsample2-D) -- node[above] {}  (Encoder_conv2-T);
	\draw [-stealth, ultra thick] (Encoder_conv2-D) -- node[above] {}  (Downsample3-T);
	\draw [-stealth, ultra thick] (Downsample3-D) -- node[above] {}  (Encoder_conv3-T);
	\draw [-stealth, ultra thick] (Encoder_conv3-D) -- node[above] {}  (Downsample4-T);
	\draw [-stealth, ultra thick] (Downsample4-D) -- node[above] {}  (Encoder_conv4-T);
	\draw [-stealth, ultra thick] (Encoder_conv4-D) -- node[above] {}  (Downsample5-T);
	
	 \path[very thick, -stealth] (Downsample5-D) edge[bend right=10]  node[above] {} (Encoder_conv5-A);
	 \path[very thick, dashed, -stealth] (Encoder_conv5-B) edge[bend right=1]  node[above] {} (Upsample1-A);
	
	\draw [-stealth, ultra thick] (Upsample1-T) -- node[above] {}  (Decoder_conv1-D);
	\draw [-stealth, ultra thick] (Decoder_conv1-T) -- node[above] {}  (Upsample2-D);
	\draw [-stealth, ultra thick] (Upsample2-T) -- node[above] {}  (Decoder_conv2-D);
	\draw [-stealth, ultra thick] (Decoder_conv2-T) -- node[above] {}  (Upsample3-D);
	\draw [-stealth, ultra thick] (Upsample3-T) -- node[above] {}  (Decoder_conv3-D);
	\draw [-stealth, ultra thick] (Decoder_conv3-T) -- node[above] {}  (Upsample4-D);
	\draw [-stealth, ultra thick] (Upsample4-T) -- node[above] {}  (Decoder_conv4-D);
	\draw [-stealth, ultra thick] (Decoder_conv4-T) -- node[above] {}  (Upsample5-D);
	\draw [-stealth, ultra thick] (Upsample5-T) -- node[above] {}  (Decoder_conv5-D);
	\draw [-stealth, ultra thick] (Decoder_conv5-T) -- node[above] {}  (Decoder_conv6-D);
	\draw [-stealth, ultra thick] (Decoder_conv6-T) -- node[above] {}  (ouput);


   \path[dashed, -stealth] (-0.1, -1.0) edge[bend left=3]  node[above] {\bf{\Large Concat.}} (13.2,-1.2);
   \path[dashed, -stealth] (-0.1, -5.0) edge[bend left=3]  node[above] {\bf{\Large Concat.}} (13.2,-5.2);
   \path[dashed, -stealth] (-0.1, -9.0) edge[bend left=3]  node[above] {\bf{\Large Concat.}} (13.2,-9.2);
   \path[dashed, -stealth] (-0.1, -13.0) edge[bend left=3]  node[above] {\bf{\Large Concat.}} (13.2,-13.2);
   \path[dashed, -stealth] (-0.1, -17.0) edge[bend left=3]  node[above] {\bf{\Large Concat.}} (13.2,-17.2);
  	

	\node[below=0.15em of O2-T] (convtext1) {\bf{\Large Conv}};
	\node[below=0.01em of Downsample1-T] (downsampletext1) {\bf{\Large Downsample}};
	\node[below=0.01em of Encoder_conv1-T] (encoderconvtext1) {\bf{\Large Conv}};
	\node[below=0.01em of Downsample2-T] (downsampletext2) {\bf{\Large Downsample}};
	\node[below=0.01em of Encoder_conv2-T] (encoderconvtext2) {\bf{\Large Conv}};
	\node[below=0.01em of Downsample3-T] (downsampletext3) {\bf{\Large Downsample}};
	\node[below=0.01em of Encoder_conv3-T] (encoderconvtext3) {\bf{\Large Conv}};
	\node[below=0.01em of Downsample4-T] (downsampletext4) {\bf{\Large Downsample}};
	\node[below=0.01em of Encoder_conv4-T] (encoderconvtext4) {\bf{\Large Conv}};
    \node[below=0.01em of Downsample5-T] (downsampletext5) {\bf{\Large Downsample}};
	\node[below=0.01em of Encoder_conv5-T] (encoderconvtext5) {\bf{\Large Conv}};
	
	\node[below=0.3em of Upsample1-T] (upsampletext1) {\bf{\Large NN Interp.}};
	\node[below=0.3em of Decoder_conv1-T] (decoderconvtext1) {\bf{\Large Conv}};
	\node[below=0.3em of Upsample2-T] (upsampletext2) {\bf{\Large NN Interp.}};
	\node[below=0.3em of Decoder_conv2-T] (decoderconvtext2) {\bf{\Large Conv}};
	\node[below=0.3em of Upsample3-T] (upsampletext3) {\bf{\Large NN Interp.}};
	\node[below=0.3em of Decoder_conv3-T] (decoderconvtext3) {\bf{\Large Conv}};
	\node[below=0.3em of Upsample4-T] (upsampletext4) {\bf{\Large NN Interp.}};
	\node[below=0.3em of Decoder_conv4-T] (decoderconvtext4) {\bf{\Large Conv}};
	\node[below=0.3em of Upsample5-T] (upsampletext5) {\bf{\Large NN Interp.}};
	\node[below=0.3em of Decoder_conv5-T] (decoderconvtext5) {\bf{\Large Conv}};
	\node[below=0.3em of Decoder_conv6-T] (decoderconvtext6) {\bf{\Large Conv + FM Avg.}};

\end{tikzpicture}}
        \vspace{1.38cm}
    \captionof{figure}{\footnotesize Model diagrams for the shared encoder and the task specific decoders. Conv refers to convolution, NN Interp. to nearest-neighbour interpolation and FM Avg. to feature map averaging.}
    \label{fig:model_details}
\end{minipage}%
\hfill
\begin{minipage}{.485\textwidth}
  \centering
    \footnotesize
    \begin{tabular}{c|c|c|c}    \hline \hline
         &Layer Type& Parameters &Array Size \\ \hline
         \multirow{ 12}{*}{$\mathcal{E}$} & Input & - & $(2, 128, 1025)$ \\
         &Conv & k(5x6), s1, p2 & $(16, 128, 1024)$ \\ \cline{2-4}
         &Downsample & k(4x4), s2, p1 & $(16, 64, 512)$ \\
         &Conv & k(3x3), s1, p1 & $(16, 64, 512)$ \\ \cline{2-4}
         &Downsample & k(4x4), s2, p1 & $(32, 32, 256)$ \\
         &Conv & k(3x3), s1, p1 & $(32, 32, 256)$ \\ \cline{2-4}
         &Downsample & k(4x4), s2, p1 & $(64, 16, 128)$ \\
         &Conv & k(3x3), s1, p1 & $(64, 16, 128)$ \\ \cline{2-4}
         &Downsample & k(4x4), s2, p1 & $(128, 8, 64)$ \\
         &Conv & k(3x3), s1, p1 & $(128, 8, 64)$ \\ \cline{2-4}
         &Downsample & k(4x4), s2, p1 & $(256, 4, 32)$ \\
         &Conv & k(3x3), s1, p1 & $(256, 4, 32)$ \\ \hline
         \multirow{ 17}{*}{$\mathcal{D}_{S_i}$}&Interpolation & NN & $(256, 8, 64)$ \\
         & Conv & k(3x3), s1, p1 & $(128, 8, 64)$ \\ 
         & Concatenate & - & $(256, 8, 64)$ \\\cline{2-4}
         & Interpolation & NN & $(256, 16, 128)$ \\
         & Conv & k(3x3), s1, p1 & $(64, 16, 128)$ \\ 
         & Concatenate & - & $(128, 16, 128)$ \\\cline{2-4}
         & Interpolation & NN & $(128, 32, 256)$ \\
         & Conv & k(3x3), s1, p1 & $(32, 32, 256)$ \\ 
         & Concatenate & - & $(64, 32, 256)$ \\\cline{2-4}
         & Interpolation & NN & $(64, 64, 512)$ \\
         & Conv & k(3x3), s1, p1 & $(16, 64, 512)$ \\ 
         & Concatenate & - & $(32, 64, 512)$ \\\cline{2-4}
         & Interpolation & NN & $(32, 128, 1024)$ \\
         & Conv & k(3x3), s1, p1 & $(16, 128, 1024)$ \\
         & Concatenate & - & $(32, 128, 1024)$ \\\cline{2-4}
         & Conv & k(3x2), s1, p1 & $(16, 128, 1025)$ \\
         & FM Avg. & k(1x1), s1, p0 & $(2, 128, 1025)$ \\
         \hline \hline
    \end{tabular}
    \captionof{table}{\footnotesize Architecture details for the shared encoder and the task specific decoders. k refers to kernel size, s to stride, p to padding, NN to nearest-neighbour and FM Avg. to feature map averaging.}
    \label{tab:architecture_details}
\end{minipage}  
\end{figure}

To train our model, we use stereo audio signals sampled at $44.1$ kHz. The STFT is computed using a window size of 2048 samples with $75$\% overlap. Patches of $L=128$ time-frames are used, resulting in input samples of size $(2, 128, 1025)$ to the network. The input magnitude spectrograms are normalized so that each frequency bin is scaled by the standard deviation computed over the whole training database. The network outputs do not undergo any kind of post-processing to refine the estimated magnitude spectrograms. This is in order to highlight the focus of this research; to verify the viability and strengths of using a shared encoder model coupled with interleaved training in a multitask setting.

\section{Training Procedure}
\label{sec:training_procedure}
Joint training of a single network to extract music sources allows the network to learn shared information across overlapping sub-tasks. With the  outlined model however,  we separate the task of learning a shared representation from the task of learning the individual specificity and unique representations embedded in each sub-task. In training such a shared encoder model, it is important to avoid a training procedure that over-fits on a sub-task. This likely results in an encoder that does not learn the shared component of the overlapping sub-tasks and consequently cannot generalize. Furthermore, it is of great importance to remove the constraint of having all target sources available for each input mixture in the database, as this allows bigger and more diverse databases.

\begin{figure}[ht]
    \centering
    \begin{tikzpicture}
    
	\node [cascadedvoice,
	fill = voicecolor,
    minimum width = 5em,
    minimum height = 5em] (VoiceDataset) {Vocals};
    
	\node [cascadeddrums,
	fill = drumscolor,
    minimum width = 5em,
    minimum height = 5em,
	right= 3em of VoiceDataset] (DrumsDataset) {Drums};
	
	\node [cascadedbass,
	fill = basscolor,
    minimum width = 5em,
    minimum height = 5em,
	right= 3em of DrumsDataset] (BassDataset) {Bass};
	
	\node [cascadedother,
	fill = othercolor,
    minimum width = 5em,
    minimum height = 5em,
	right= 3em of BassDataset] (OtherDataset) {Other};
    
    \node [cascadeddd,
	fill = othercolor,
	draw= othercolordark,
    minimum width = 5em,
    minimum height = 5em, below= 7.5em of BassDataset.south west] (InterleavedDataset) {Interleaved Batch};
    
    \draw[very thick, lines, dashed] (VoiceDataset.south east) -- ([xshift=-3em, yshift=3.5em]InterleavedDataset.north west);
    \draw[very thick, lines, dashed] (DrumsDataset) -- ([yshift=2.5em]InterleavedDataset.north west);
    \draw[very thick, lines, dashed] (BassDataset) -- ([xshift=-1em, yshift=1.35em]InterleavedDataset.north);
    \draw[very thick, lines, dashed] (OtherDataset) -- ([xshift=2em, yshift=.35em]InterleavedDataset.north);
    
\end{tikzpicture}
    \caption{\footnotesize Interleaved training. Combining mini-batches from the Vocals, Drums, Bass and Other independent databases to train our shared encoder model.}
    \label{fig:interleaved_training}
\end{figure}
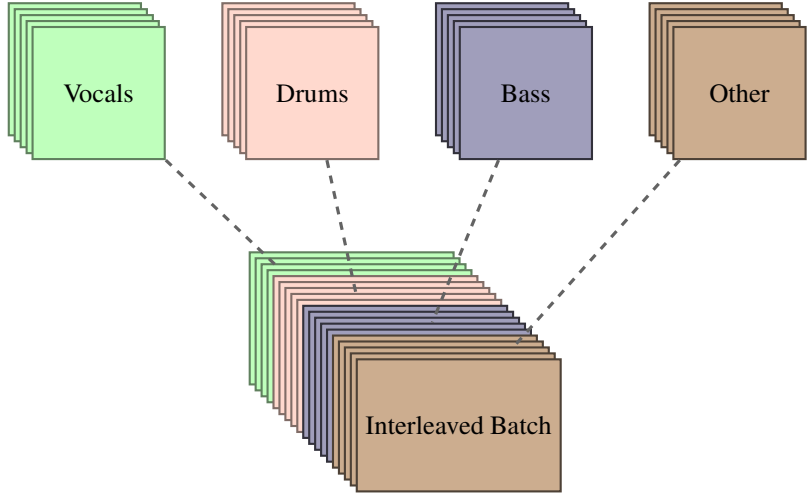

The training procedure implemented combines the individual sub-task databases, effectively yielding an interleaved training procedure. Thus, our model is trained by alternating mini-batches of data corresponding to the sub-tasks, as demonstrated in Figure~\ref{fig:interleaved_training}.

\subsection{Interleaved training}
\label{ssec:interleaved}

The proposed training procedure interleaves the sub-task specific databases in a non-repeating task order. Where sub-task databases are of unequal length, random subsets equal in length to the smallest sub-task database are used. These randomly sampled subsets vary across epochs. Subsequently, the encoder is trained by alternating the sub-task mini-batches of samples passed. The layers of the encoder are chained to the corresponding decoder, whose loss optimizes both decoder and encoder as per the sub-task. A detailed description of this training procedure is shown in Algorithm~\ref{alg:interleaved} for one epoch, where $B$ is the number of batches available in the smallest sub-task database and the loss for each sub-task mini-batch is computed according to (\ref{eq:l1_loss}).

The interleaving progressively alternates the sub-task the encoder is optimized for and thus constrains the encoder to learn a joint representation of the multitask space which is useful to all sub-tasks. The increased variation in input mixture samples also prevents over-fitting to individual sub-tasks. The combined effect is a disentangling between the common distribution of the overlapping sub-spaces, and the task specific distribution functions.

\subsection{Accumulating gradients at the encoder}
\label{ssec:gradient_accumulation}

We also implement a variation of the interleaved training procedure, where the back-propagated gradients are accumulated on the encoder for the whole interleaved batch. This means that the encoder weights $\theta_{\mathcal{E}}$ are updated after a batch from all the underlying sub-tasks has been seen, whereas the decoder weights $\theta_{\mathcal{D}_{S_i}}$ are updated after each sub-task mini-batch. The procedure is detailed in Algorithm~\ref{alg:accumulated} for one epoch.

We hypothesize that updating the encoder after each sub-task mini-batch induces a stochastic behaviour, as each update might not be in the direction of a weight configuration that best characterizes the overlapping distribution. By summing the gradients, we attempt to introduce update normalization.

\begin{minipage}{.48\textwidth}
\hrulefill
\vspace{0.1cm}
\begin{algorithmic}
\For{$b = 1, ..., B$}
        \For{$\bm{S}_{i} \in \mathcal{S}$}
            \State \text{Sample mini-batch of data pairs $\left(\bm{Y}_{S_i}, \bm{S}_i\right)$}
            \State $\hat{\bm{S}}_i = \mathcal{D}_{S_i}\left( \mathcal{E}\left(\bm{Y}_{S_i};\theta_E\right);\theta_{\mathcal{D}_{Si}}\right)$
            \State $\theta_{\mathcal{E}} \leftarrow \theta_{\mathcal{E}} - \lambda \nabla_{\theta_{\mathcal{E}}}\mathcal{L}_{S_i}$
            \State $\theta_{\mathcal{D}_{S_i}} \leftarrow \theta_{\mathcal{D}_{S_i}} - \lambda \nabla_{\theta_{\mathcal{D}_{S_i}}}\mathcal{L}_{S_i}$
        \EndFor
\EndFor
{} \\
{} \\
\end{algorithmic}
\vspace{0.1cm}
\hrulefill
\captionof{algorithm}{Interleaved training}\label{alg:interleaved}
\end{minipage}
\hfill
\begin{minipage}{.48\textwidth}
\hrulefill
\vspace{0.1cm}
\begin{algorithmic}
\For{$b = 1, ..., B$}
        \State $G_{\mathcal{E}} = 0$
        \For{$\bm{S}_{i} \in \mathcal{S}$}
            \State \text{Sample mini-batch of data pairs $\left(\bm{Y}_{S_i}, \bm{S}_i\right)$}
            \State $\hat{\bm{S}}_i = \mathcal{D}_{S_i}\left( \mathcal{E}\left(\bm{Y}_{S_i};\theta_E\right);\theta_{\mathcal{D}_{Si}}\right)$
            \State $G_{\mathcal{E}} \, \pluseq \nabla_{\theta_{\mathcal{E}}}\mathcal{L}_{S_i}$
            \State $\theta_{\mathcal{D}_{S_i}} \leftarrow \theta_{\mathcal{D}_{S_i}} - \lambda \nabla_{\theta_{\mathcal{D}_{S_i}}}\mathcal{L}_{S_i}$
        \EndFor
        \State $\theta_{\mathcal{E}} \leftarrow \theta_{\mathcal{E}} - \lambda G_{\mathcal{E}}$
\EndFor
\end{algorithmic}
\vspace{0.1cm}
\hrulefill
\captionof{algorithm}{Accumulated-encoder interleaved training}\label{alg:accumulated}
\end{minipage}

\section{Evaluation}
\label{sec:evaluation}

\subsection{Evaluation metrics}
\label{ssec:metrics}

To evaluate our source separation model, we use the Blind Source Separation (BSS) metrics; Source-to-Distortion Ratio (SDR) and Source-to-Interferences Ratio (SIR) \cite{vincent2006}. Many papers only look at the SDR to compare between algorithms \cite{takahashi2017, park2018}, as it is a more general measure of performance encompassing all possible errors in the output under the distortion term. However, we believe the SIR, measuring the ratio between the true source component and the projection of the other sources in the model's output, is very relevant perceptually. Furthermore, because the encoder is shared in our model's architecture, the SIR can be viewed in this case as a measure of how well each decoder succeeded in disentangling the task-specific distribution functions. Finally, we do not include the Source-to-Artifacts Ratio (SAR) scores as we found them to be of limited interest in this particular case. Indeed, because the interference term at the denominator of the SIR goes to the numerator of the SAR, we found the latter to follow more or less the opposite trend of SIR scores in most cases.

Following the procedure used in the SiSec 2018 separation campaign \cite{SiSEC18}, the BSS metric values were computed using the BSSEval v4 implementation on non-overlapping segments of one second. As detailed in \cite{stoller2018}, near-silent segments can be problematic and lead to strong outliers. Thus, we took the median SDR and SIR values over all the segments within each song, and computed the average of the resulting song scores across the test database.

\subsection{Implementation details}
\label{ssec:implem_details}

All trainings were performed with the ADAM optimizer \cite{kingma2014}, employing a learning rate of 0.0002 and a batch size of 32 samples per GPU (or the maximum possible if 32 samples could not fit in memory). Training was automatically stopped when no improvement was observed on the validation set in the preceding 30 epochs, and the learned parameters leading to the  lowest validation loss were selected.

All of the outlined models were trained using the MUSDB18 database \cite{SiSEC18}. The MUSDB18 training set contains $100$ song samples, which we randomly split into $75$ samples for training and $15$ samples for validation. All models were evaluated using the test set of MUSDB18. Data augmentation was used to amplify the variance of these samples. The augmentations used were: random scaling, panning and filtering of segments of the individual sources. 

In order to evaluate the impact of using the interleaved training procedure with independent databases while still using the same amount of data, two versions of the training set were generated. A \emph{simultaneous database}, where for each input mixture the corresponding four sources are set up as targets, and four \emph{independent databases}, one for each target source, where a different random training/validation split is used for each source, and the order of the data pairs (input mixture, target source) is shuffled. This is projected to simulate the case of unavailability of all target sources for any given input mixture.

\subsection{Impact of the training procedure}
\label{ssec:training}

We aim to observe the effect and impact of training a shared task model using the proposed interleaved procedures. To observe this effect, we keep the model as constant as possible during experimentation but vary the training procedure. Since the focus of the present work is on sharing the encoder between tasks and can be transposed to any network architecture, we did not compare it against other baseline architectures.

Four training procedures are compared for the base encoder and decoder architectures described in Section~\ref{ssec:architecture_details}. We denote these four training procedures as: \begin{itemize}[leftmargin=*]
    \item {\bf Independent} a separate network is trained for each of the target objectives using the independent databases. It is worth noting that since the encoder is not shared between tasks in this case, the networks inherently have more capacity. This serves as a baseline.
    \item {\bf Simultaneous} a shared encoder and four independent decoders trained with the four target objectives optimized simultaneously using $\mathcal{L} = \sum _{S_{i} \in \mathcal{S}} \mathcal{L}_{S_i}$. Trained with the simultaneous database.
    \item {\bf Interleaved} a shared encoder and four independent decoders trained with the interleaved method described in Section~\ref{ssec:interleaved}. Trained with the independent databases.
    \item {\bf Interleaved}$_{\bf acc}$ a shared encoder and four independent decoders trained with the accumulated encoder gradients interleaved method described in Section~\ref{ssec:gradient_accumulation}. Trained with the independent databases.
\end{itemize}

\begin{table}[ht]
    \centering
    \setlength\tabcolsep{5pt}
    \begin{tabular}{c|c c c c|c c c c}
    \hline \hline
         Training & \multicolumn{4}{c|}{SDR in dB} & \multicolumn{4}{c}{SIR in dB} \\
         Method & Vocals & Drums & Bass & Other & Vocals & Drums & Bass & Other \\ \hline
         Independent & 4.15 & 4.74 & 3.11 & 2.38 & 9.51 & 7.37 & 4.01 & 2.18\\
         Simultaneous & 3.35 & 4.64 & 2.96 & 2.49 & 7.62 & 6.74 & 3.65 & 2.48\\
         Interleaved & 3.55 & 4.54 & 2.67 & 2.28 & 8.92 & 7.25 & 3.8 & 2.04\\
         Interleaved$_{acc}$ & 3.33 & 4.61 & 2.61 & 2.36 & 8.02 & 7.37 & 2.9 & 1.89\\ \hline \hline
    \end{tabular}
    \vspace{0.1cm}
    \caption{Average song SDR and SIR scores in dB on the test set of MUSDB18.} 
    \label{tab:sdr_training}
\end{table}

As can be seen in Table~\ref{tab:sdr_training}, we observe that learning independent models for each sub-task leads to higher SDR scores than using a multitask representation. This observation can be made for all sub-tasks and all multitask training methods except one: the highest SDR score for the separation of the "other" stem is obtained through simultaneous training. We believe this can be explained by the fact that separating vocals, drums or bass from a mixture signal are well defined tasks and therefore easier to learn with dedicated networks. On the other hand, identifying the source which does \emph{not} contain vocals, drums or bass is a very ill-defined and inconsistent task for an independent network. Indeed, the latter is presented with vastly different targets for each input mixture, and would need to learn implicitly which sources to remove. There is a tendency in that case to learn a model with poor generalization.

The shared encoder networks trained with either interleaved procedure do not perform as well as independent networks in terms of SDR scores, most notably for separating the vocals, and perform similarly to the simultaneously trained shared-encoder network. Based on the definition of the SDR metric, low values can be interpreted as a deficiency in learning the individual representation of each source. This can be attributed to a propensity to get stuck in a local minimum due to the complexity of the shared cost space. We also see that the interleaved training with accumulation of the encoder gradients leads to similar performance for both drums and bass objectives, slightly better performance for other, and worse performance in vocals separation than the non-accumulated version.

The SIR measures the ability of the model to disentangle sources, especially in the case of a shared encoder where it implicitly learns a joint representation on a set of related sub-tasks. It can be seen in Table~\ref{tab:sdr_training} that interleaved training leads to significantly higher SIR scores than simultaneous training. This shows its ability to learn a shared representation and use that representation in the encoder to inform the disentangling of sources in the individual decoders. However, this is true for all sources except "other" where, as was the case with SDR scores, simultaneous training leads to the best performance among all training methods. Finally, the independently trained networks show better SIR scores on the vocals and bass separation task, with the non-accumulated interleaved procedure being a close second. We believe the accumulated version of the interleaved training normalizes gradient directions through accumulation and thus loses the stochasticity which might help in navigating through complex shared cost spaces. Thus it performs worse than its non-accumulated counterpart.

\cite{Caruana1997} hypothesized that the reduced model capacity induced by the shared encoder leads to better generalization through implicit regularization. The results outlined above indicate that for well defined tasks such as separating vocals, drums or bass this might not be the case, as model capacity seems to be a very important factor. The proposed interleaved training procedure manages to mitigate the effects of the reduced model capacity, getting comparable SIR results to independent networks with less than half the parameters ($\approx 8$M (million) for the 4 separate networks together against $\approx 3.5$M for the shared encoder architecture). It is also worth noting that these experiments are performed with an equal amount of data between all training procedures, and therefore do not take advantage of the fact that the independent databases used in the independent and both interleaved cases could be greatly enhanced.

\subsection{Increasing the encoder capacity}
\label{ssec:encoder_capacity}

In order to put the shared encoder architecture and the independent networks on more equal terms with respect to the model capacity, we use the interleaved training procedure of Section~\ref{ssec:interleaved}, while increasing the width (i.e. the number of computed feature maps) of the encoder network. 

The model denoted by "encoder" implies the architecture remains unchanged from Section ~\ref{ssec:architecture_details} and therefore the results are those reported in Table~\ref{tab:sdr_training}. The same encoder architecture with slightly more computed feature maps at the higher resolution layers (i.e. 32 feature maps computed instead of 16 for the first 3 layers) is referenced as "encoder+". In both cases, the number of computed feature maps at the decoders remains unchanged from that described in Section~\ref{ssec:architecture_details}. This means the architecture denoted by encoder+ has approximately $50\,000$ more parameters to train on top of the base $3.5$M.

\begin{table}[ht]
    \centering
    \setlength\tabcolsep{5pt}
    \begin{tabular}{l|c c c c|c c c c }
    \hline \hline
         & \multicolumn{4}{c|}{SDR in dB} & \multicolumn{4}{c}{SIR in dB} \\
          & Vocals & Drums & Bass & Other & Vocals & Drums & Bass & Other \\ \hline
         encoder & 3.55 & 4.54 & 2.67 & 2.28 & 8.92 & 7.25 & 3.8 & 2.04 \\
         encoder+ & 3.61 & 4.79 & 2.88 & 2.58 & 9.28 & 7.43 & 4.37 & 2.25 \\ \hline \hline
    \end{tabular}
    \vspace{0.1cm}
    \caption{Average song SDR score in dB on the test set of MUSDB18. Training procedure and architecture are constant, only the number of computed feature maps at the encoder changes.}
    \label{tab:sdr_width}
\end{table}

It can be observed for encoder+ in Table~\ref{tab:sdr_width} that the slight increase in the number of computed feature maps at the topmost layers leads to an increase in SDR scores across all target source objectives, the effect being stronger for the "other" objective and smaller for the vocals objective. The SIR scores shown in Table~\ref{tab:sdr_width} show a very similar trend to their SDR counterpart. The results are even more pronounced as encoder+ shows higher SIR scores than independently trained networks for all target sources but the vocals, for which the results are slightly lower. 

We believe that increasing the number of computed feature maps at the topmost layers increases the variance of the extracted features and this, in turn, allows the network to make use of a wider range of possible features. Thus, in a network constrained by expressive power but with a large feature base, the centering of the learned overlapping distribution might be tilted towards the sub-task(s) requiring the largest amount of available features, or the least defined sub-task(s).

These results show that slightly increasing the encoder capacity at higher resolutions allows to close the performance gap with independently trained networks, and even go further, while having far fewer trainable parameters, more efficient inference, and keeping the advantage of being able to use independent databases. Finally, the learned shared encoder representation is inherently transferable to new source separation objectives (e.g. guitar, piano), where it could also be used as an excellent starting point for training new source-specific networks.

\section{Conclusion}
\label{sec:conclusion}

In this paper, we presented a multitask learning approach to the audio source separation problem as well as a training procedure that interleaves the optimization of each sub-task, allowing the use of independent databases. Using a shared-encoder architecture with increased width at the highest layers, we achieve state-of-the-art performance compared to separate networks trained independently on each task, with fewer parameters and more efficient inference.

\bibliographystyle{unsrt}  


\end{document}